# Dynamics of the electro-optic response of blue bronze $K_{0.3}MoO_3$

L. Ladino, J.W. Brill, M. Freamat,[a] M. Uddin, and D. Dominko[b]

Department of Physics and Astronomy, University of Kentucky, Lexington, KY 40506

(Dated: May 19, 2006)

Abstract

We have studied the charge density wave (CDW) repolarization dynamics in blue bronze by applying square-wave voltages of different frequencies to the sample and measuring the changes in infrared transmittance, proportional to CDW strain. The frequency dependence of the electro-transmittance was fit to a modified harmonic oscillator response and the evolution of the parameters as functions of voltage, position, and temperature are discussed. Resonant frequencies decrease with distance from the current contacts, indicating that the resulting delays are intrinsic to the CDW with the strain effectively flowing from the contact. For a fixed position, the average relaxation time has a voltage dependence given by $\tau_0 \sim V^{-p}$, with $1<p<2$. The temperature dependence of the fitting parameters shows that the dynamics are governed by both the force on the CDW and the CDW current: for a given force and position, both the relaxation and delay times are inversely proportional to the CDW current as temperature is varied. The long relaxation and delay times ($\sim 1$ ms) suggest that the strain response involves the motion of macroscopic objects, presumably CDW phase dislocation lines.

I. Introduction

Interest in quasi-one dimensional conductors with sliding charge-density-waves (CDW's) has continued for three decades because of the variety of unusual properties they exhibit.[1,2] In the CDW ground state, a periodic lattice distortion is accompanied by a modulated electron density: $n = n_0 + n_1 \cos[Qz + \varphi(z,t)]$, where z is the direction of the conducting chains and Q is the CDW wavevector, $n_1$ its amplitude, and $\varphi$ its local phase. Because CDW pinning results from the deformation of the CDW (i.e. variations of $n_1$ and $\varphi$) due to its interaction energy with impurities, CDW's are also model systems for studying the effects of quenched disorder on a deformable periodic medium.[3] This competition results in the CDW responding to different stimuli at a very large array of time scales.[3,4]

When a voltage greater than its depinning threshold is applied, the CDW can slide through the sample, carrying current.[1] At the same time, it becomes elastically strained, i.e. its phase varies throughout the sample so that the CDW is compressed near one current contact and rarefied at the other.[3,5-8] The strain ($Q^{-1}\partial\varphi/\partial z$) near a current contact is required to drive the phase-slip process needed for current conversion, i.e. to allow electrons to enter and leave the CDW condensate at the contacts.[6,9] Strains near the center of the sample reflect shifts in the chemical potential due to imperfect screening by "normal" quasiparticles, i.e. those from non-condensing bands and/or those thermally excited out of the CDW.[7] The spatial-temporal dependence of the strain in turn affects that of the CDW current.[6,7]

Consequently, the illumination of the spatial dependence of the CDW current and strain, through tunneling,[10] transport,[6,11,12] x-ray,[7,13] and infrared measurements,[8] has been a major thrust area for the last several years. The Cornell group addressed this in an elegant series of conductivity experiments on NbSe$_3$.[6,12] By closely spacing non-perturbative contacts along the sample, they measured local electric fields and solved model transport equations to deduce the local strain as functions of position and time after *reversing* the direction of current flow. Their results indicated that strains near (~ 100 μm) the current contacts reversed quickly (~ 10 μs) whereas the smaller strains in the center took several times longer to change. Subsequently, spatially resolved x-ray diffraction measurements at Grenoble[7] directly measured the CDW strains in NbSe$_3$ and observed much larger contact deformations than those deduced by the Cornell group. The difference was attributed to the fact that in the Cornell model[6] the quasiparticle conductivity was treated as a constant, whereas in the model of Reference [7] the uncondensed and thermally excited quasiparticles respond differently to CDW strains, resulting in local changes in their density, larger strains near the contacts, and more extended regions of current conversion. The strains near the center of the sample persisted even when the applied voltage was removed;[5,7,13] these non-equilibrating strains were associated with pinning of CDW phase dislocations,[7] discussed further below.

Whereas NbSe$_3$ remains metallic in the CDW state, in most other sliding CDW materials all conduction bands are gapped in the CDW state.[1] The CDW's in these semiconducting materials are much less coherent than in NbSe$_3$, preventing x-ray measurements of the spatial dependence of the strain. Analysis of position dependent transport properties of these materials has also been complicated by the difficulty in preparing Ohmic but non-perturbative contacts.[11] The absence of uncondensed electrons furthermore complicates the analysis of transport properties because as the density of thermally excited quasiparticles falls at low temperatures, soliton-like defects can dominate even the low-field, Ohmic, conductance.[11]

Instead, our group has used measurements of infrared transmittance and reflectance to measure position and time dependent changes in the semiconducting CDW phase in $TaS_3$[4] and $K_{0.3}MoO_3$ (blue bronze).[8,14,15] The infrared changes occur primarily because, even at relatively high temperatures, the quasiparticle density is low enough that CDW strains can cause relatively large changes in their density, resulting in changes in intraband absorption.[8,15,16] In addition, changes in phonon frequencies and widths have been observed.[16-18] At any wavelength, the relative changes in transmittance, $\Delta\theta/\theta$, and reflectance, $\Delta R/R$ are assumed to be essentially proportional to the local CDW strain.[8,16]

In our previous measurements on blue bronze, we found that $\Delta\theta/\theta$ varied approximately linearly with position in the sample for small applied voltages.[8] The onset voltage for the electro-optical effect, $V_{on}$, typically occurred below the threshold, $V_T$, for non-Ohmic resistance (i.e. non-linear current), as shown for a present sample in Figure 1. The difference, $V_T-V_{on}$, was associated with the "phase-slip voltage[9]" in our 2-probe measurements, for which the same contacts on the ends of the sample are used as current and voltage leads; that is, $V_{on}$ was associated with the onset of CDW depinning and strain in the sample bulk whereas $V_T$ was associated with depinning at the contacts, allowing CDW current to flow. For voltages above $V_T$, additional small changes in transmittance, which we associated with the extra strains needed for phase-slip,[6] were observed near (within ~ 100 μm of) the current contacts.[8] As in Figure 1, most of these measurements were made with symmetric square-wave voltages applied to the sample, so that the local strain oscillated between opposite signs, and our ac measurements were sensitive to the resulting oscillating transmitted IR signal. We also found that the signal associated with the "bulk", linear strain decreased with increasing square wave frequency, with a characteristic frequency (~ 0.1 - 1 kHz) which decreased with decreasing temperature, while the strains near the contact were faster, but we did not do detailed measurements of the spatial/frequency dependence at that time.[8,15]

If unipolar square-waves, in which the voltage oscillated between zero and a voltage of one sign, were applied, no oscillating signal was generally observed in the center of the sample, and only small ac signals associated with the extra contact strains, were observed.[14,17] That the contact strain oscillates with turning the voltage on and off again indicates its connection to current conversion and phase-slip. The bulk strain, on the other hand, was apparently frozen when the applied voltage was removed, consistent with the persistent strains observed with transport[5] and x-ray[7,13] measurements. An important difference between our blue bronze observations and the x-ray results on $NbSe_3$,[7] however, is that in our case the observed contact signal is much smaller than the persistent bulk signal. While this may be a consequence of the difficulty in preparing non-perturbative Ohmic contacts in blue bronze, as discussed below, it may also reflect the fact that the greater CDW incoherence in blue bronze, as compared to $NbSe_3$, makes the bulk strain, associated with poor screening and possibly dislocation line pinning, more dominant.

In this paper, we report on detailed measurements of the frequency dependence of the electro-transmittance of blue bronze as functions of voltage, position, and temperature. We recently reported on similar measurements in $TaS_3$,[4] but for that material, the poor sample (and CDW) morphology made position dependent measurements difficult to interpret. A short discussion of some of the present results was given in Reference [19].

## II. Experimental details

Blue bronze is a quasi-one dimensional metal with a transition at $T_c$ = 180 K into a semiconducting CDW state.[1,20,21] As grown blue bronze crystals were cleaved with tape to thicknesses of 5±2 μm for these transmittance measurements. Gold or copper contacts (~ 500 Å) were evaporated on the ends of the samples so that the resulting lengths between the contacts, in the high-conductivity direction, were 0.5 – 1.0 mm, while the sample widths were typically ~ 0.2 mm. Fine silver wires glued to the contacts with conducting paint acted as both flexible current leads and thermal grounds for the fragile samples which were mounted over ~ 1mm holes in sapphire substrates in our vacuum cryostats. The sample temperature was determined by a thermometer placed close to the sample, but the sample temperatures may have been a few degrees warmer because of the incident IR light and Joule heating from the applied currents.

Infrared light from a tunable infrared diode laser was focused on the sample using an infrared microscope. The light, polarized along the sample width and with multimode power ~ 0.1 mW, was incident (on the face of the sample containing the contacts) through a rectangular aperture into a spot 50 μm along the length of the sample and 80 μm perpendicular. The light position was measured between the closest edges of the light spot and a metal film contact (x=0), but the contact edges were not always sharp, so the absolute positions are uncertain by ~ 20 μm; in addition, the aperture image typically drifted with respect to the sample during a measurement by ~ 10 μm.

For relative changes in transmittance or reflectance, the infrared beam was chopped (at 390 Hz) and a square-wave voltage applied to the sample. The modulated transmitted or reflected signal was measured simultaneously at the square-wave ($\omega$) and chopping frequencies; the ratio of the signals precisely gives $\Delta\theta/\theta$ or $\Delta R/R$, even though the transmittance or reflectance were not separately measured precisely.[17,18] Most measurements were made with symmetric bipolar square-waves, for which the changes in transmittance and reflectance are $\Delta\theta \equiv \theta(+V) - \theta(-V)$ and $\Delta R \equiv R(+V) - R(-V)$. Some unipolar transmittance measurements, for which $\Delta\theta \equiv \theta(V) - \theta(0)$, were also made. In both cases, the responses both in-phase and in quadrature with the square-wave were measured; the frequency-dependent phase shift of the microscope detector electronics was determined with a precision of 2°, possibly slightly affecting results at the lowest frequencies, where the quadrature component becomes much smaller than the in-phase component.[4]

Thin samples were used to help insure uniform current flow through the sample cross-section. This was checked by comparing the dependence of the electro-transmittance and electro-reflectance on position, voltage, and frequency, as shown in Figure 2. As seen, only the absolute magnitudes of $\Delta\theta/\theta$ and $\Delta R/R$ differed, with $\Delta\theta/\theta \sim 3 \Delta R/R$, but they had the same spatial, voltage, and frequency dependences, indicating that the CDW phase gradient and current were uniform across the sample cross-section, at least on the scale of the penetration depth of the light, typically a few times smaller than the sample thickness.[16,17] Detailed measurements were then made primarily on the transmittance, for which the microscope was less susceptible to mechanical noise than reflectance. For each sample, a wavelength was chosen for which the laser power and electro-transmittance were large. (For a sample with $\alpha d > 1$, where $\alpha$ is the absorption coefficient and d the sample thickness, the transmittance is given by $\theta \sim (1-R)^2 \exp(-\alpha d)$, making the electro-transmittance spectrum sample thickness dependent.) The photon energies used were between 775 and 890 cm$^{-1}$, on a broad plateau in the $\Delta\theta/\theta$ spectrum associated with quasiparticle absorption.[17]

Samples were also chosen for having fairly regular, linear variations of $\Delta\theta/\theta$ with position when measured with bipolar square-waves. An example is shown in Figure 3. Samples with irregular spatial dependences presumably contained macroscopic defects (e.g. cracks, grain boundaries) that affected the CDW current flow but were not apparent from visible inspection. An interesting feature of Figure 3, that we observed in most samples, is that the zero crossing of $\Delta\theta/\theta$, i.e. the position of zero CDW strain, varies with voltage. As discussed further in Section VI, this suggests that the relative "quality" of the two contacts varies somewhat with voltage.

Also shown in Figure 3 is the spatial dependence when a unipolar square-wave is applied. (Note that the magnitude of the unipolar response is roughly consistent with the deviation from linear spatial dependence near the contact observed for the bipolar response at the same voltage.) As discussed above, the transmittance will only oscillate for applied unipolar voltages for strains which decay when the voltage is removed, and that these "non-pinned" strains were previously only observed to occur adjacent to the contact.[14,17] Now, we observe that there are also small unipolar variations in the center of the sample; similar observations were made for other (but not all -- see Figure 12, below) samples. The unipolar spatial dependence shown in Figure 3 suggests that the non-pinned strain may "overshoot", flowing into adjacent regions and leaving them with strains of opposite sign during the "on" half cycle and/or strains of the same sign in the "off" half cycle.

All the present measurements are at temperatures a) well above the "CDW-glass" transition (23 K) where the dielectric time constant diverges[22] and high enough that b) the low-field resistance has activation energy ~ half the gap, indicating that the Ohmic transport is dominated by thermally activated quasiparticles,[11] and c) the non-Ohmic conductivity is dominated by thermal activation over pinning barriers and only exhibits a single threshold voltage without hysteretic switching.[23]

### III. Frequency dependence of bipolar response at T = 80 K

We have measured in detail the voltage and position dependence of the frequency dependence (2 Hz $< \omega/2\pi <$ 4 kHz) of the electro-transmittance at T ~ 80 K for three samples, of lengths L = 550 μm (sample 1), L = 810 μm (sample 2) and L = 990 μm (sample 3), using bipolar square-waves. For bipolar square-waves, the response contains both the bulk, non-equilibrating portion of the strain which would stay pinned at V = 0 and the non-pinned, contact strain. For samples 1 and 2, however, the non-pinned component piece was extremely small (e.g. see Figure 3). It was relatively larger for sample 3, but as discussed in Section IV, still did not affect the fits significantly. We therefore assume that the dynamics of the bipolar response is always essentially that of the non-equilibrating, bulk strain.

Representative data sets at x = 0 (adjacent to a contact) and x = 200 μm for the three samples are shown in Figures 4, 5, and 6. The following features, all qualitatively similar to TaS$_3$,[4] can be seen: a) At x = 0, the response is essentially relaxational. There is a peak in the quadrature response at the same frequency at which the in-phase response falls; the peak height is roughly half the magnitude of the low frequency in-phase response.   b) The relaxation peak moves to lower frequencies with decreasing voltage. c) At x = 200 μm, the response is smaller and slower than at x = 0 for each voltage.  d) At high frequencies, the in-phase response becomes inverted, corresponding to a *delay* in the electro-optical response.  This occurs at much lower frequencies for x = 200 μm than at x = 0.  (It is most noticeable at x = 0 for sample 1; for sample 3 it is generally out of our frequency window.)   e) For some voltages and positions, the quadrature signal becomes inverted at low frequencies.  It is clearest for sample 1, where it occurs in some cases for frequencies > 50 Hz.  For the other two samples, it only occurs for $\omega/2\pi$ < 10 Hz, where noise and the difficulty in determining the electronics phase shift makes this inversion less certain.

The relaxation and delay, and its strong position dependence, can also be seen in time-averaged oscilloscope traces of the transmitted signal (with the chopper turned off), as shown in Figure 4c, where for sample 1 a delay of ~ 70 μs is observed at x = 100 μm and V = 3.6 $V_{on}$, but the delay is only ~ 20 μs for the same voltage at x =0.  As for TaS$_3$,[4] the delayed reversal of transmittance seems to begin "abruptly", especially at x =0.  We do not observe the striking polarity dependence of the delay observed for TaS$_3$, however.  Also, the delay time increases much more rapidly as one moves away from the contacts for blue bronze than for TaS$_3$, indicating that the delay is *not,* as we suggested in Reference [4], a contact effect, e.g. due to the formation of Schottky barriers at the contacts,[11] but is intrinsic to the pinning and repolarization of the CDW.

As shown in Figures 4, 5, and 6, we have characterized the relaxation and delay by fitting the complex response to the modified harmonic-oscillator equation[4]

$$\Delta\theta/\theta = \Delta\theta/\theta)_0 / [1 - (\omega/\omega_0)^2 + (-i\omega\tau_0)^\gamma]. \qquad (1)$$

Here the inertial term is used to parameterize the delays. The exponent $\gamma$ allows for distributions in $\omega_0$ and $\tau_0$; in particular, for overdamped cases $\gamma < 1$ corresponds to a distribution of relaxation times.[24] The low-frequency inverted quadrature signal for sample 1, discussed further below, is not included in these fits.

The voltage and position dependence of the fitting parameters are shown in Figure 7. For each sample, the relaxation times increase slightly as one moves away from the contact and for each sample/position, the relaxation time is seen to vary as $\tau_0 \sim V^{-p}$, with p between 1 and 2, as for $TaS_3$.[4] (We note that while a similar value of $\tau \sim 1$ ms at $V_T$ was deduced for the rate of polarization from an unpolarized state, it had a much stronger, exponential, voltage dependence.[25]) Note that the onset of CDW current at $V_T$ appears to have no effect on the relaxation time. The resonant frequencies tend to weakly ($\sim V^{1/2}$) increase with increasing voltage, but much more striking is the rapid increase in resonant frequency (i.e. decrease in inertia) as one approaches the contact, with the resonant frequency in some cases exceeding our frequency window. The amplitudes decrease rapidly as one approaches $V_{on}$; away from the contacts, the amplitudes start decreasing again at high voltage whereas they approximately saturate at x =0.[8] Finally, for samples 2 and 3 the exponent $\gamma$ decreases at small voltages as the resonance becomes increasingly overdamped, indicating that the distribution of relaxation times is becoming broad. For example, $\gamma \sim 0.9$ corresponds to a distribution of $\tau$'s approximately a half decade wide while $\gamma \sim 2/3$ corresponds to a distribution over a decade wide.[24] (This broadening may mask the expected dynamic critical slowing down expected at $V_{on}$.[26])

At any given voltage, or depinning force $(V-V_{on})/L$ or $(V-V_T)/L$, sample 2 is much slower than samples 1 or 3; e.g. its relaxation times are over an order of magnitude larger and resonant frequencies an order of magnitude lower. (The time constants for samples 1 and 3 are close to those of $TaS_3$.[4]) As we'll discuss in Section IV, this suggests that sample 2 is effectively "colder" than the other samples so that the CDW current density is less for a given driving force. Our preliminary data on sample 2 suggests that this is so but, unfortunately, the sample broke before measurements could be made at different temperatures or larger currents. (Note that the threshold and onset electric fields for sample 2 are close to that of the shorter sample 1 but much greater than those of sample 3.)

The inverted quadrature signal observed at low frequencies, especially for sample 1, corresponds to a long-time *decay* of the electro-optic response. (The small magnitude of this decay and poor signal/noise ratio of low-frequency oscilloscope measurements prevented us from directly observing this decay in a time trace like that of Figure 4c.) Since $\gamma \sim 1$ for sample 1, the response can be fit by including a term $i\Omega_x^2 \tau_0/\omega$ in the denominator of Eqtn. 1, where $\Omega_x$ is the frequency at which the quadrature signal changes sign. However, $\Omega_x$ had no clear position or voltage dependence, so the fits are not shown. Furthermore, the significance of the decay is not clear. It should be emphasized that this decay is observed for bipolar square waves, including cases for which $|V|$ is always above threshold, so is not due to the slow decay of bulk strains at $V=0$.[7] In the Cornell model, the elastic force on the CDW (the gradient of the strain) does decay with time but no decay is expected for the strain itself.[6]

### IV. Temperature dependence of time constants

The dynamics of many properties associated with CDW pinning/depinning are governed by the density of quasiparticles,[3] and hence time constants are thermally activated for semiconducting materials. The dielectric constant, associated with small oscillations of the CDW about its pinning configuration, is a very prominent example.[22,27] The magnitude of strain for a given voltage, since it depends on screening by quasiparticles, is also expected to be activated, and indeed the magnitude of the electro-transmittance was observed to be activated at low temperatures.[15] On the other hand, in the Cornell model, the rate of change of strain, when reversing the current, is governed not only by the magnitude of the strain (near the contacts) but by the CDW current.[6,12] It was therefore interesting to study the dynamics of the electro-transmittance at different temperatures to see how these different factors are manifested.

Measurements were made on sample 3 for temperatures between 45 K and 100 K. (The decreasing resistance prohibited measurements at voltages much above threshold at higher temperatures, while the decreasing magnitude of the electro-transmittance prevented frequency dependent measurements at lower temperatures.) To make comparisons of the response at different temperatures, we had to decide on an appropriate voltage criterion. Both the onset and threshold voltages[14,21] decrease with decreasing temperature in this range, so temperature-dependent comparisons at a constant voltage are not appropriate. ($V_T/V_{on}$ increases with decreasing temperature[14] so that most of the pinning becomes associated with the contacts, e.g. see Figure 1. However, $V_T$-$V_{on}$ actually decreases from 8 mV to 3 mV with decreasing temperature for this sample, in striking contrast with suggested thermal activation of the phase-slip voltage.[28]) Figure 7 also shows that the relaxation time does not change as the voltage passes through $V_T$ (and hence must also have a weak dependence on CDW current here). Therefore, we chose as our criterion fixed values of V-$V_{on}$, approximately proportional to the average driving force on the strain and CDW. (The pinning force, in fact, depends on the voltage and CDW current for small currents.[6])

Figure 8 shows the temperature dependence of the parameters of Eqtn. 1 for two driving voltages, V = $V_{on}$ + 10 mV and V = $V_{on}$ + 20 mV, and two positions, x = 0 and x = 200 μm. Also shown is the temperature dependence of the low-field, quasi-particle resistance and conductance. For each position/voltage, the magnitude of the electro-transmittance varies roughly as the conductance for T < 70 K and saturates at higher temperature, as we previously observed.[15] In contrast, for each data set the relaxation times and resonant frequencies have a weaker temperature dependence, with activation energies approximately half that of the conductance. However, the CDW current for each voltage, defined as $I_{CDW} \equiv I_{TOTAL} - V/R_0$, was observed to also have this smaller activation energy. We therefore plotted the relaxation times and frequencies vs. $I_{CDW}$, as shown in Figure 9. ($I_{CDW}$ is plotted vs. voltage in the inset to Figure 9; for this sample, $I_{CDW} \sim (V-V_T)^{3/2}$.) For each voltage/position, $\tau_0$ and $1/\omega_0$ scale roughly as $1/I_{CDW}$; that is, *for a given driving force and position, the dynamical rates are proportional to the CDW current*.

We should emphasize that this dependence on CDW current only occurs for a given driving voltage; the relaxation rates themselves are not simply functions of $I_{CDW}(T)$, even for large CDW currents. This is shown in Figure 10, where we plot the fit parameters of Eqtn. 1 for fixed $I_{CDW}$ = 100 μA. While the temperature dependence of the rates are weaker than for fixed driving voltage, they are still strongly temperature dependent.

To check this dependence of the dynamical rates on driving voltage and CDW current, we did temperature dependent measurements on an additional sample. For sample 4 (L ~ 960 μm), the fits to Eqtn. 1 were not as good as for the other samples, especially at the highest and lowest frequencies, but the fits at intermediate frequencies were sufficient to determine relaxation times. As shown in the inset to Figure 11, the relaxation times for this sample tended to saturate at low temperatures, very different behavior from sample 3. Nonetheless, as shown in Figure 11, there is still a linear dependence, for each voltage/position, between the relaxation rate and CDW current.

### V. Frequency dependence of contact strains

For most samples, the contact strain measured with unipolar square-waves above threshold has been too small for frequency dependent measurements. An exception was sample 3, for which the spatial dependence of the unipolar (V = $2V_T$) and bipolar (V = 1.5 $V_T$) responses at T = 101 K, $\omega/2\pi$ = 25 Hz are shown in the inset to Figure 12. For this sample, the unipolar response disappears away (> 200 μm) from the contacts.

Figure 12 shows the frequency dependence of the bipolar response and both the positive and negative unipolar responses at 101 K, $2V_T$, and x=0. The parameters for the fits to Eqtn. 1 are listed in Table I. Note the following features: i) The negative unipolar response is larger than the positive, as we previously observed.[14] Polarity dependent strains have also been observed in NbSe$_3$ and are not understood.[7,12] ii) Pronounced relaxation peaks in the quadrature response are not observed for unipolar excitation. In fact, the unipolar response can be fit to Cole-Cole relaxation,[29] i.e. Eqtn. 1 without the resonance term, as shown in the Table, but with a very small $\gamma=0.65$, implying a decade wide distribution in relaxation times,[24] which effectively washes out the relaxation peak. iii) Given this wide distribution in $\tau$'s, the differences in relaxation times between the bipolar and unipolar responses is probably not significant. Indeed, it is striking that the "non-pinned", contact unipolar response is not much faster than the (mostly non-equilibrating) bipolar response. This suggests that changes in the contact strains, as the CDW current is turned on and off, occur through a similar mechanism as oscillations in the bulk strain, as discussed below.

Also shown in Figure 12 is the bipolar response at $x = 200$ μm, where the unipolar response is zero. As usual, the striking difference between the response here and at the contact is that the resonant frequency has moved well into our window; in fact the resonance at $x = 200$ μm is underdamped even though the relaxation time (see Table I) has also increased considerably. In Reference [8], we speculated that the faster bipolar response at the contact vs. the interior might be due to the fact that at the contact the bipolar response comes from both non-pinned and non-equilibrating strains, whereas the response in the interior only comes from non-equilibrating strains. However, the difference between the bipolar and two unipolar responses, also shown in Figure 12, gives the $x=0$ non-equilibrating response only. This difference was also fit to Eqtn. 1 and its parameters listed in Table I. The difference response is only slightly slower (i.e. relaxation time is essentially the same and resonant frequency only slightly smaller) than that of the full bipolar response, showing that the dynamics of the bipolar response is essentially determined by that of the non-equilibrating part of the strain, even at points adjacent to the contact.

## VI. Summary and discussion

We have used measurements of the changes in infrared transmittance when square-wave voltages are applied to the sample to determine the position, voltage, and frequency dependence of CDW strains (i.e. phase gradients) in blue bronze. This technique has the advantage of being able to probe the interior of the sample without placing multiple probes, which can perturb the CDW, on the sample, which has hindered transport measurements of the CDW strain in semiconducting CDW materials.[11]

Of course, it is still necessary to place contacts on the ends of the sample, and the largest changes in dynamical properties occur near (~ 100 μm) these contacts. Important questions, therefore, are to what extent these contacts are equipotentials with minimal band-bending, e.g. due to formation of Schottky barriers in the CDW state,[11] and to what extent current enters the sample from the edges of the contacts and quickly distributes through the cross-section. The very small contact strains observed for some samples and the fact that the bulk strain, while varying approximately linearly with position in the sample, is not symmetric on the two sides of the sample with a voltage dependent asymmetry (e.g. see Figure 3) certainly suggest that our contacts are "imperfect". Indeed, the longitudinal length-scale with which current is expected to spread through the sample cross-section from surface contacts is[11] $\lambda \sim d\, \eta^{1/2}$, where $\eta$ is the ratio of longitudinal and transverse conductivities. For blue bronze, $\eta \sim 1000$,[20] making $\lambda$ comparable to the length scale of measured changes at the contacts.

However, the fact that the electro-transmittance (probing the whole sample cross-section) and the electro-reflectance (probing only the < 2 µm penetration depth[16,17]) have the same spatial and frequency dependence (Figure 2) suggests that current spreading is not a significant problem for our contacts. (Note that the spatial dependence of the electro-reflectance is different for thicker samples.[18]) Two possible reasons are that sample defects effectively distribute the current through the cross-section in a distance shorter than $\lambda$ or that, because the contacts are over 100 µm long, the current actually spreads below them. The rapid variation of our measured relaxation times and resonant frequencies near the contacts suggests that the first effect is dominant. In fact, the expected length scale for the contact strain is determined by the single particle diffusion length and is expected to be ~ 100 µm,[7] consistent with our measurements. We therefore assume that CDW current is approximately injected from the edges of our contacts so that the relatively small contact strains we measure are not artifacts of poor contacts but intrinsic, e.g. due to the incoherence of the CDW and/or strong pinning of dislocations, discussed below.

Most of our measurements were for the oscillating response to symmetric bipolar square-wave voltages, so that the CDW strain is oscillating between two opposite configurations. In this case, the oscillating strain has contributions from both the bulk polarization, which does not decay in zero field, and a non-pinned strain associated with current conversion, but the latter is small and does not significantly affect the overall frequency dependence.

For large voltages, the electro-optic response can be fit as a damped harmonic oscillator, with the resonant frequency corresponding to a delay with respect to the applied square-wave voltage. The delay times increase rapidly between x = 0 (the contact) and x = 100 μm and then continue increasing (by 50-100% ) between x = 100 and 200 μm. This spatial variation indicates that the delays are intrinsic to the CDW (i.e. not associated with contact barriers[4]) and suggests that the signal driving the strain relaxation effectively flows out of the contacts. In this case, the delays we observe near the contacts may be a consequence of our finite spatial resolution. Similarly, in their measurements on NbSe$_3$, the Cornell group found that there was a delay ~ 10 μs for changes in the electric field, and therefore the strain, in the center of the sample, but no delay at the contacts, and simulated these effects in terms of a strongly strain (and therefore position and voltage) dependent phase-slip rate, presumably reflecting the pinning and motion of dislocation lines.[6] In our case, the resulting typical (T = 80 K) propagation velocity of ~ 100 μm/100 μs is comparable to that observed for voltage pulses[30] but orders of magnitude larger than the drift velocity of the CDW,[21] whose motion is limited by scattering with quasiparticles.[1] If we model the CDW strain propagation as a wave on a stretched wire with tension (per electron) ~ e(V-V$_{on}$)/L ~ 10 eV/m (consistent with the observed $\omega_0$ ~ V$^{1/2}$ behavior at large voltages), then a velocity of 1 m/s corresponds to a reasonable effective mass density for the strain wave of ~ 1 QM$_F$, where Q is the CDW wavevector and M$_F$ is the Fröhlich mass (~ 300 m$_e$[31]) associated with CDW motion.[1] Of course, this simple result should only be considered order of magnitude and needs to be qualified to account for the strong sample and temperature dependence of the delays.

The relaxation time also increases as one moves away from the current contact. For any position in the sample, the average relaxation time $\tau_0 \sim V^{-p}$, with p between 1 and 2. No divergence or other structure is observed in $\tau_0$ near $V_T$, where the CDW is depinned at the contacts and dc CDW current can flow. The dynamics are governed both by the force on the CDW (e.g. $V-V_{on}$) and the CDW current. For a given force and position in the sample, both the relaxation and delay times are inversely proportional to the CDW current as temperature is varied. The temperature dependent screening of the quasiparticles directly affects the amplitude of the electro-optic response,[15,16] but it only appears to affect the dynamics through its influence on the CDW current, so that while the response slows with decreasing temperature, it does not slow as much as expected from the quasiparticle density.

Near $V_T$ at T ~ 80K, the typical relaxation time is 1-10 ms, more than three orders of magnitude greater than the dielectric response time governing small amplitude oscillations of the pinned CDW,[22,27] indicating that repolarization requires large scale rearrangements of the CDW and suggesting the strain relaxation involves the motion of extended defects in the CDW. In the model of Reference [6], CDW polarization occurs essentially through continuous and gradual changes in CDW phase. However, this may be a coarse-grained average of a more complicated phase landscape in which regions in which CDW has its equilibrium wavevector are separated by localized, soliton-like defects.[3,32,33] These presumably accumulate on neighboring chains to form extended defects, such as CDW phase dislocation loops.[32,33,34] As mentioned above, in Reference [7] the pinning of the dislocations was considered to be the cause of persistence of the bulk strains when V=0.  If so, polarization current would require the lateral motion of these defects along the conducting chains, e.g. glide of the dislocations,[33] in contrast to the growth of the loops by climb perpendicular to the chains, responsible for phase-slip.[28,32,34]  As mentioned above, a similar mechanism would hold for the changes in the contact strains, but the wider distribution of relaxation times might indicate a broader distribution of dislocation loop sizes near the contacts where phase-slip is also occurring.

For low voltages, the simple damped harmonica fits break down (for most samples), and the response requires a distribution of time constants, which we have parameterized with the exponent $\gamma$.  This may reflect inhomogenous CDW pinning (and current, when above $V_T$) on a length scale much smaller than our typical 50 μm light spot.  As mentioned above, this broadening may mask any possible dynamic critical slowing down at $V_{on}$.[26]  However, sample 1 did not exhibit this broadening, at least for V ~ $V_T$., and it showed some evidence of critical behavior at lower voltages; unfortunately, the sample broke before the low voltage range could be investigated well.  Since the presence of dynamic critical behavior at depinning has long been an open question that has generated much interest,[35] we will continue measurements on crystals at low voltages; our electro-optic probe may allow us to avoid some of the problems encountered with transport measurements.[11,35]

In conclusion, we have used infrared electro-transmittance to probe the dynamics of CDW repolarization.  Long (millisecond) relaxation and delay times suggest that the response involves the motion of macroscopic objects, presumably CDW phase dislocation lines;[7] this appears to be true for the contact strains driving phase-slip as well as the bulk polarization.  Most striking is the growth in the delay time with position away from the contacts.  The temperature dependence of the time constants suggests that quasiparticles influence the dynamics of polarization only through their effect on the CDW current.  In future work, we will attempt to generalize the models of References [6,7] to accommodate semiconducting materials, with temperature and position dependent quasiparticle densities, to more quantitatively account for some of these observations.

We appreciate helpful discussions with S.N. Artemenko (Russian Academy of Sciences) and R.E. Thorne (Cornell University), who also kindly provided crystals.  This research was supported by the National Science Foundation, Grant DMR-0400938.

**Table I.** Comparison of Eqtn. 1 fit parameters (see Figure 12) for different square-waves with $V = 2V_T = 3.1V_{on}$ for *sample 3* at T = 101 K.

| square-wave/position | $\Delta\theta/\theta)_0$ | $\tau_0$ (ms) | $\omega_0/2\pi$ (kHz) | $\gamma$ |
|---|---|---|---|---|
| **Bipolar, x=0** | 1.05% | 0.15 | 4.70 | 0.98 |
| **+ unipolar, x=0** | 0.15% | 0.053 | $\infty$ | 0.65 |
| **- unipolar, x=0** | -0.23% | 0.13 | $\infty$ | 0.65 |
| **Bipolar, x = 200 μm** | 0.49% | 0.46 | 0.55 | 0.94 |
| **Bip. – [(+uni.) – (-uni.)], x=0** | 0.70% | 0.16 | 2.92 | 1.06 |

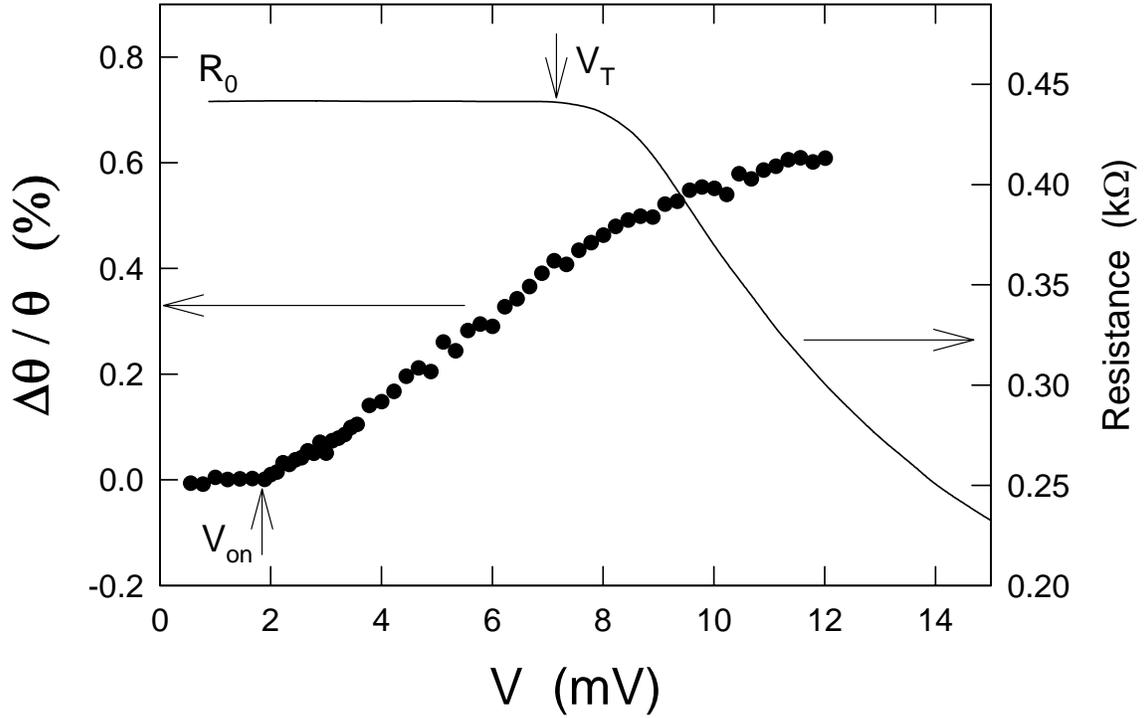

**Figure 1.** Dependence of the dc resistance and relative change in transmittance (ν = 820 cm$^{-1}$) for *sample 3* at T ~ 80 K on voltage across the sample. The transmission is measured at a point adjacent to a current contact (x =0) in-phase with a bipolar square-wave at 25 Hz, for which the quadrature changes are negligible. The low-field "ohmic" resistance associated with quasiparticle current ($R_0$), threshold voltage for non-linear current ($V_T$) and onset voltage for the electro-optic response ($V_{on}$) are indicated.

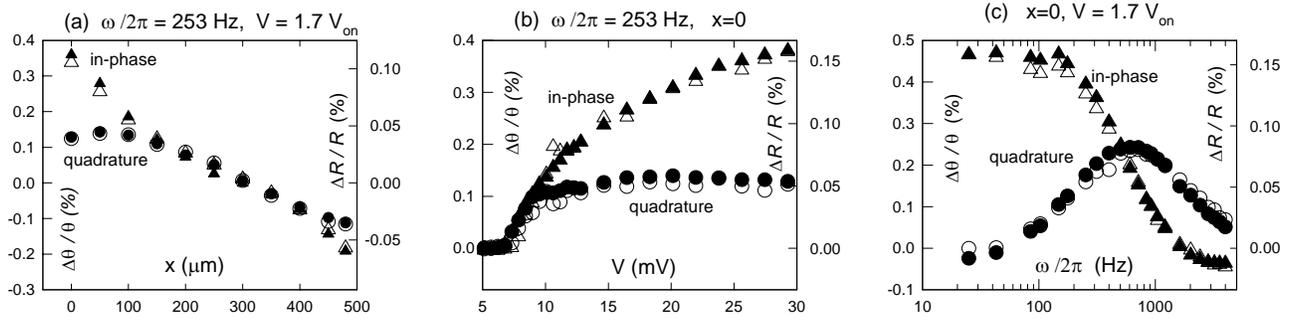

**Figure 2.** Comparison of the (a) spatial, (b) voltage, and (c) frequency dependences of the electro-transmittance (solid symbols, $\nu = 820$ cm$^{-1}$) and electro-reflectance (open symbols, $\nu = 850$ cm$^{-1}$) for *sample 1* at T ~ 80 K at the frequencies, voltages, and positions indicated. Both the response in-phase and in quadrature with the driving bipolar square-waves are shown.

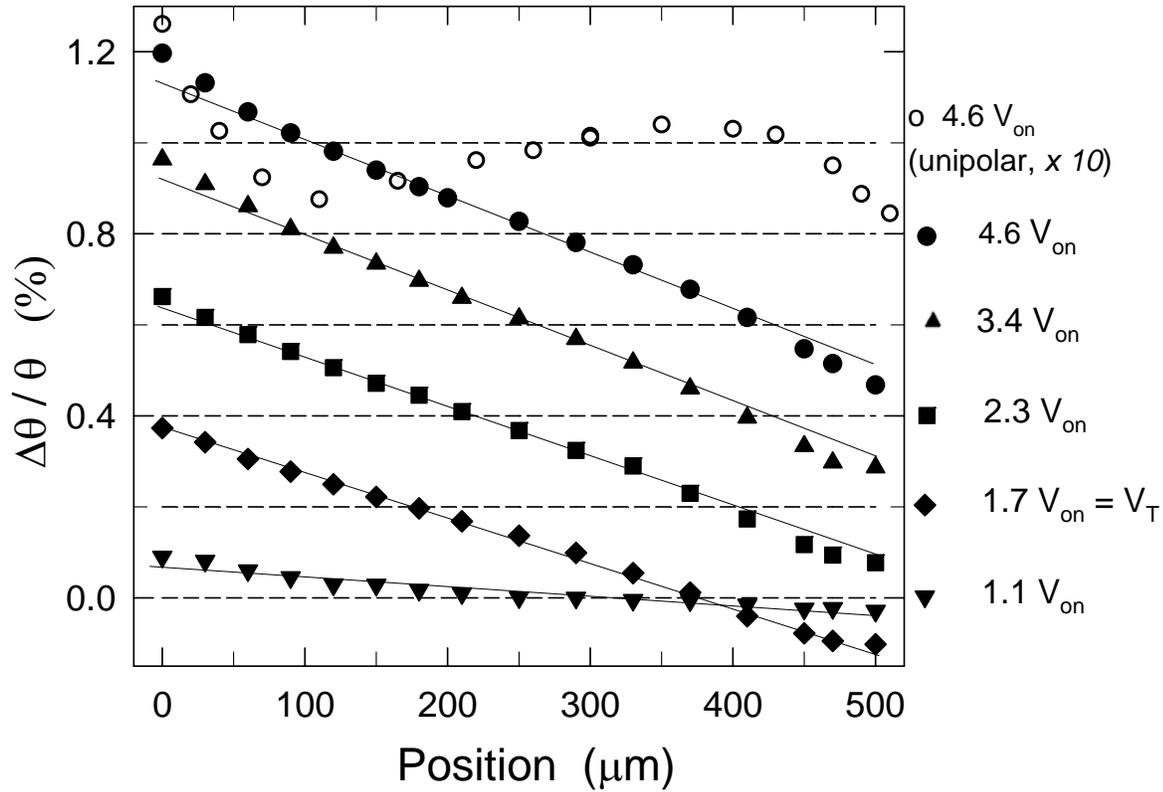

**Figure 3.** The spatial dependence of the electro-transmittance ($\nu = 820$ cm$^{-1}$) of *sample 1* at T ~ 80 K in-phase with bipolar square waves at several voltages at 25 Hz, for which the quadrature response is negligible. The sample was 550 μm long and the light spot was 50 μm wide. Each data set is vertically offset by 0.2%; the dashed zero-line for each data set is shown, with the voltage given on the right. The solid lines through the data points are for reference only. The open symbols show the response for a positive unipolar square wave (*multiplied by 10*).

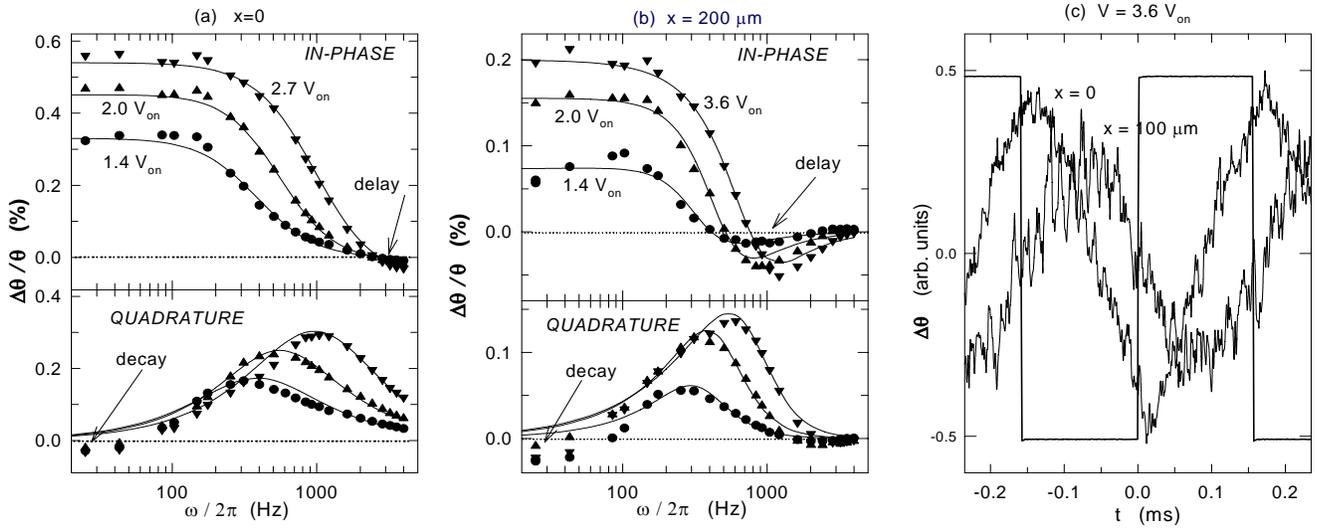

**Figure 4.** (a,b) The frequency dependence of the electro-transmittance ($\nu = 820$ cm$^{-1}$) of *sample 1* at a few bipolar square-wave voltages at a position (a) adjacent to a current contact and (b) 200 μm from the contact; both the response in-phase and in quadrature with the square wave are shown. The arrows indicate the high-frequency inverted in-phase response associated with delay and the low-frequency inverted quadrature response associated with long-time decay of the electro-optical signal. The curves are fits to Eqtn. 1. (c) Electro-transmittance vs. time for $V = 3.6$ V$_{on}$, $\omega/2\pi = 3.2$ kHz bipolar square wave at two positions; the applied square-wave is shown for reference.

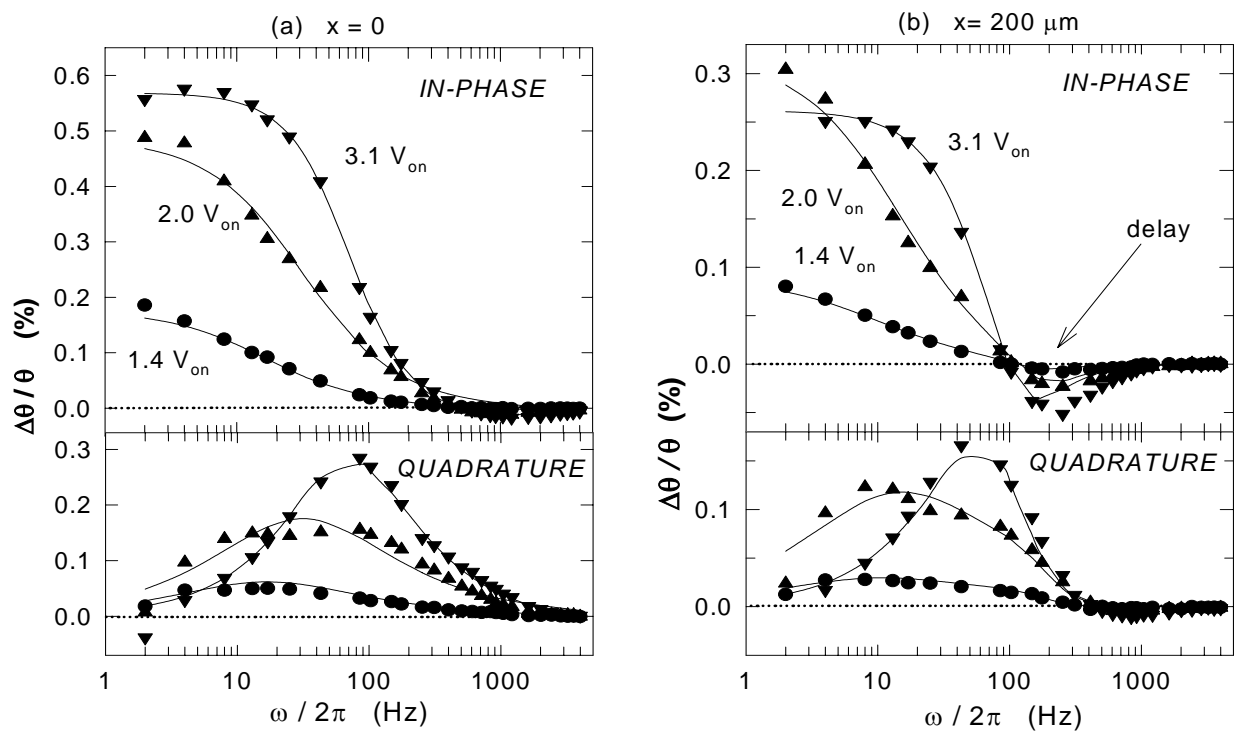

**Figure 5.** The frequency dependence of the electro-transmittance ($\nu = 890$ cm$^{-1}$) of *sample 2* at a few bipolar square-wave voltages at a position (a) adjacent to a current contact and (b) 200 μm from the contact; both the in-phase and quadratures responses are shown. The arrow indicates the high-frequency inverted in-phase response associated with delay for x = 200 μm. The curves are fits to Eqtn. 1.

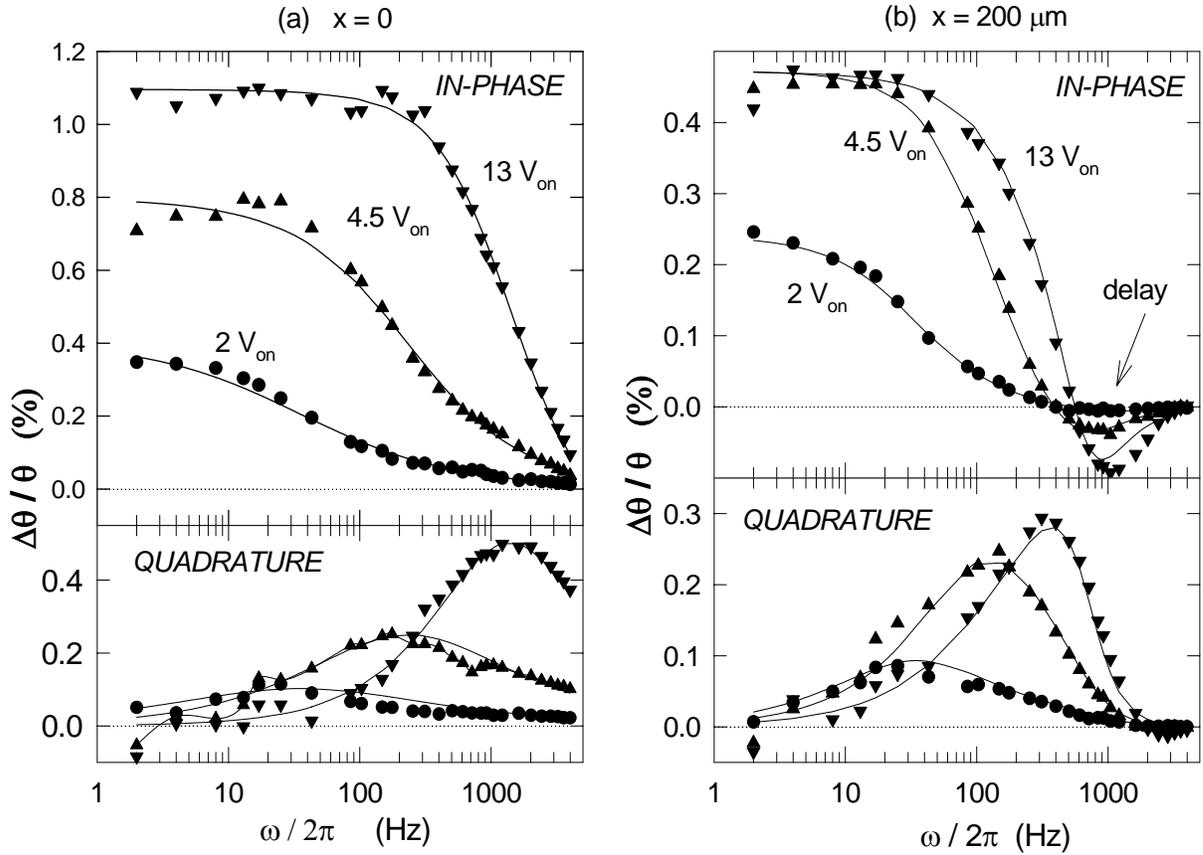

**Figure 6.** The frequency dependence of the electro-transmittance ($\nu = 820$ cm$^{-1}$) of *sample 3* at a few bipolar square-wave voltages at a position (a) adjacent to a current contact and (b) 200 μm from the contact; both the in-phase and in quadrature responses are shown. The arrow indicates the high-frequency inverted in-phase response associated with delay for x = 200 μm. The curves are fits to Eqtn. 1.

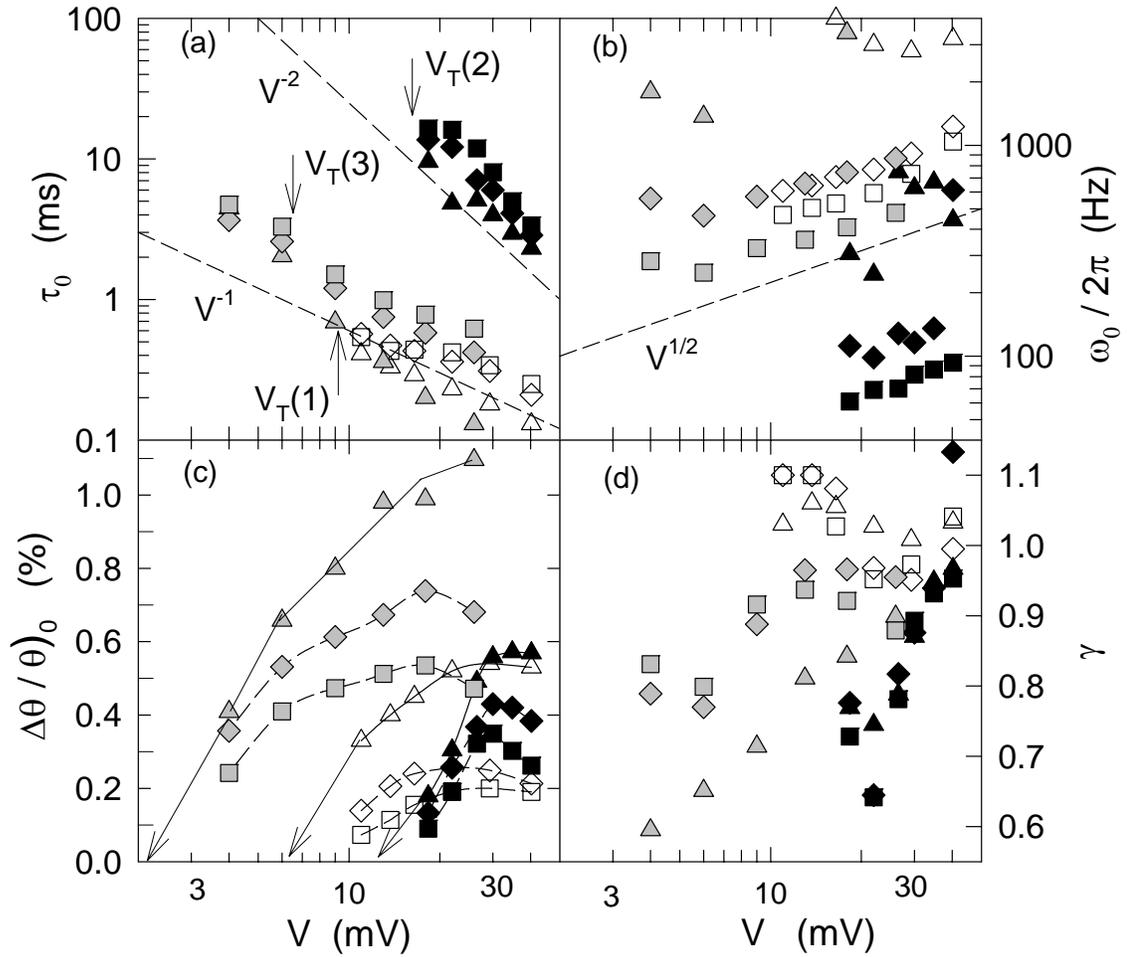

**Figure 7.** Voltage dependence of fitting parameters for Eqtn. 1 for *sample 1* (open symbols), *sample 2* (black symbols), and *sample 3* (grey symbols) at T ~ 80 K: a) relaxation times, b) resonant frequencies, c) amplitudes, d) exponents. The triangles are for x=0, diamonds for x= 100 μm, and squares for x = 200 μm. Reference lines showing $1/V$ and $1/V^2$ behavior are shown in (a) and $V^{1/2}$ behavior in (b). The vertical arrows in (a) indicate the non-linear current threshold voltages. In (c), the curves are guides to the eye, with extrapolated arrows showing the onset voltages.

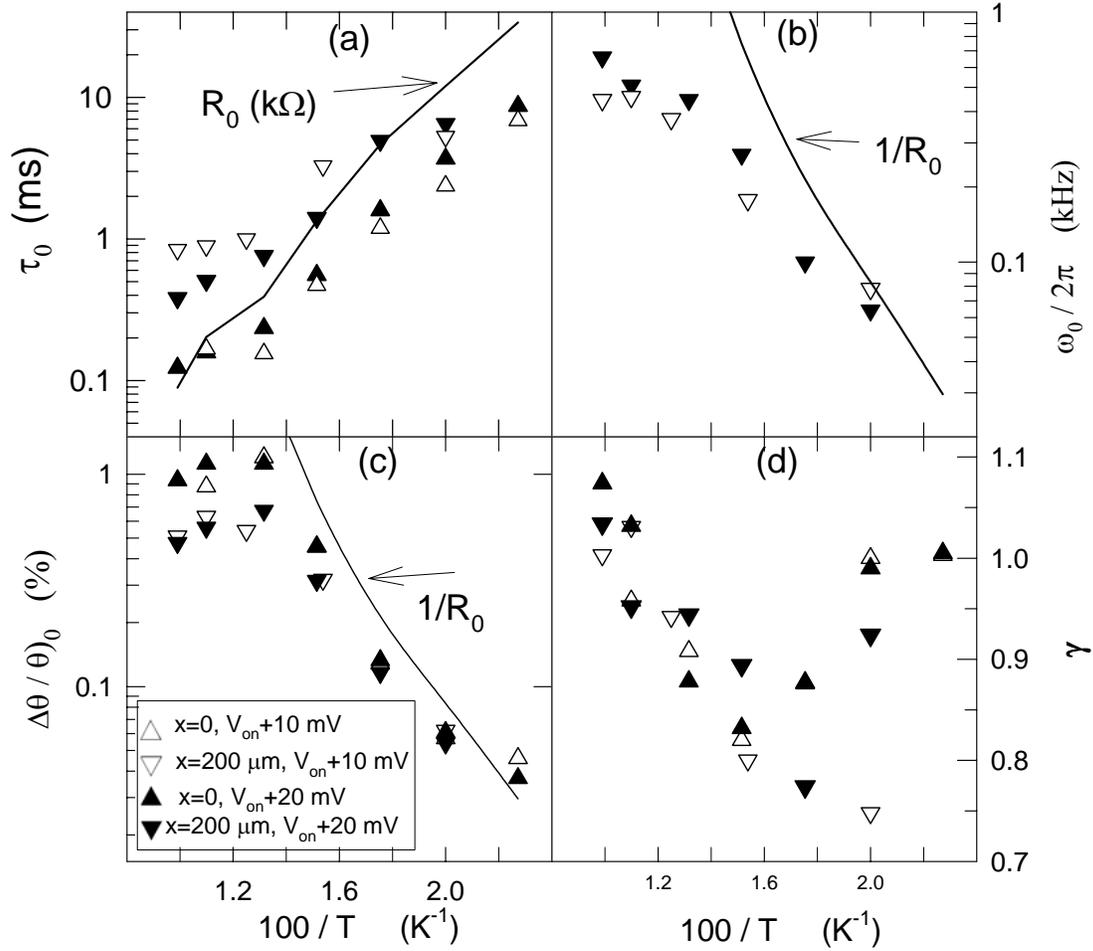

**Figure 8.** Temperature dependence of frequency-dependence fitting parameters for Eqtn. 1 for *sample 3* ($\nu = 820$ cm$^{-1}$) for two positions and voltages, as indicated: a) relaxation times, b) resonant frequencies (only determined for $x = 200$ μm), c) amplitudes, d) exponents. The curves in (a,b,c) show the temperature dependence of the low-field resistance and conductance.

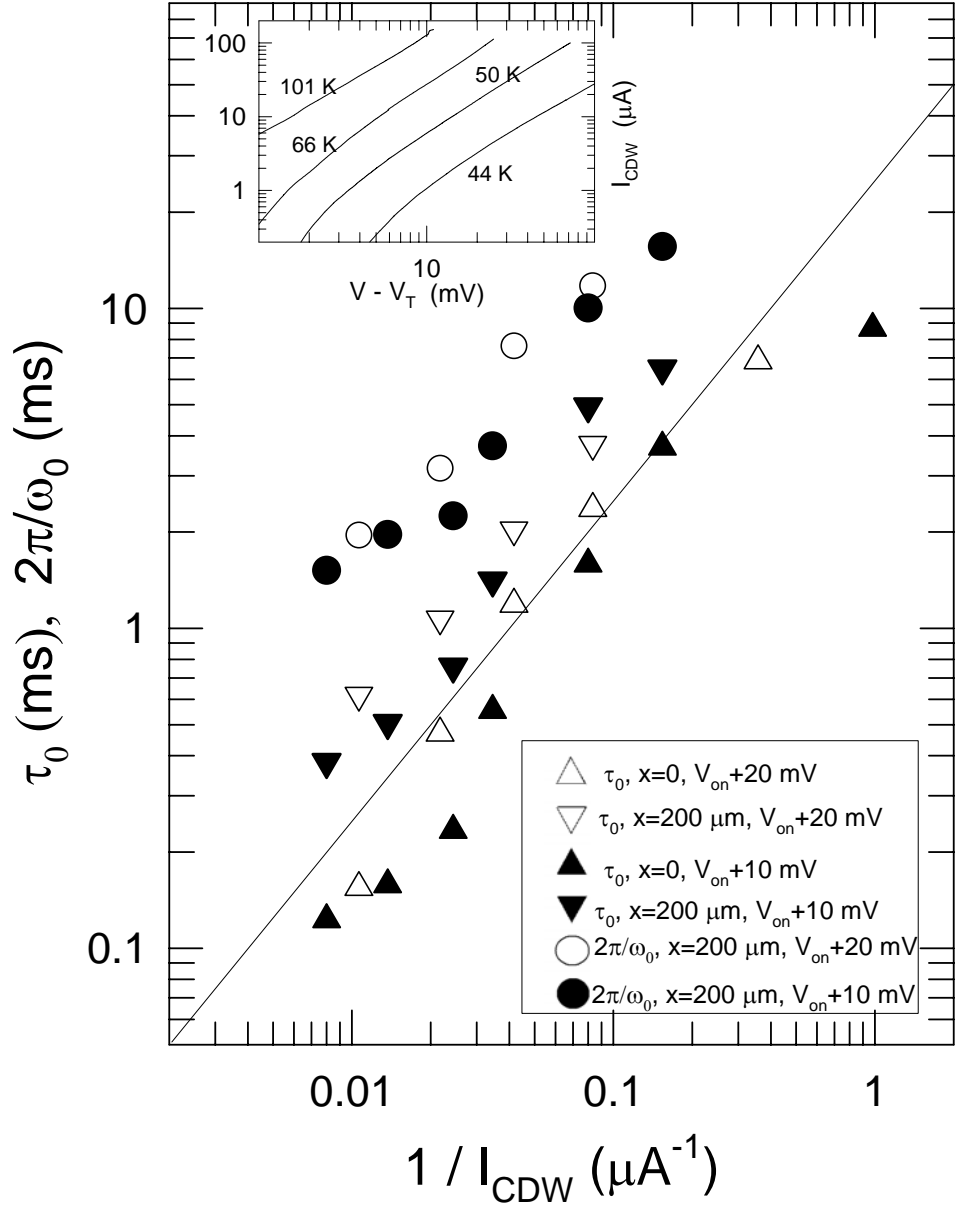

**Figure 9.** Relaxation times and resonant frequencies for *sample 3* ($\nu = 820$ cm$^{-1}$) plotted as functions of the temperature dependent CDW current at $V = V_{on} + 10$ mV and $V_{on} + 20$ mV. A line with slope = 1 is shown for reference. *Inset:* CDW current vs. voltage at a few temperatures.

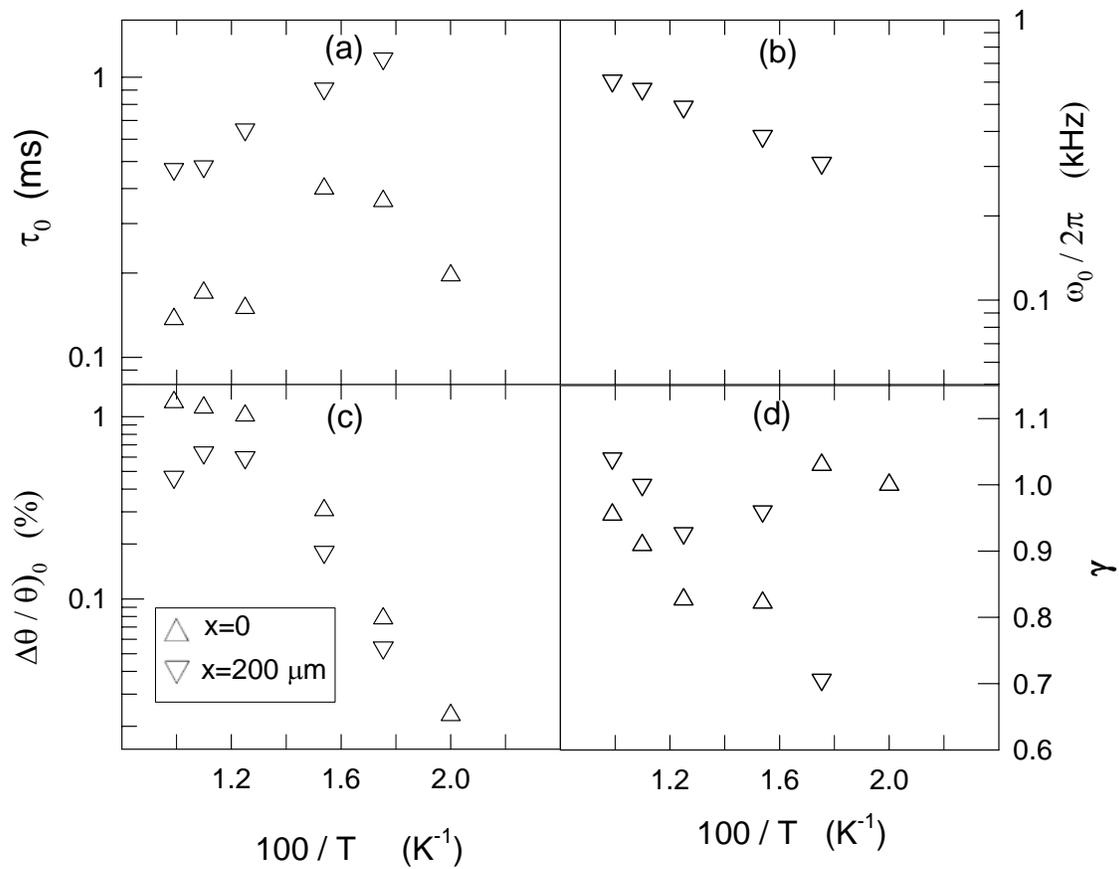

**Figure 10.** Temperature dependence of frequency-dependence fitting parameters for Eqtn. 1 for *sample 3* ($\nu = 820$ cm$^{-1}$) for $I_{CDW} = 100$ μA and two positions: a) relaxation times, b) resonant frequencies (only determined for x = 200 μm), c) amplitudes, d) exponents.

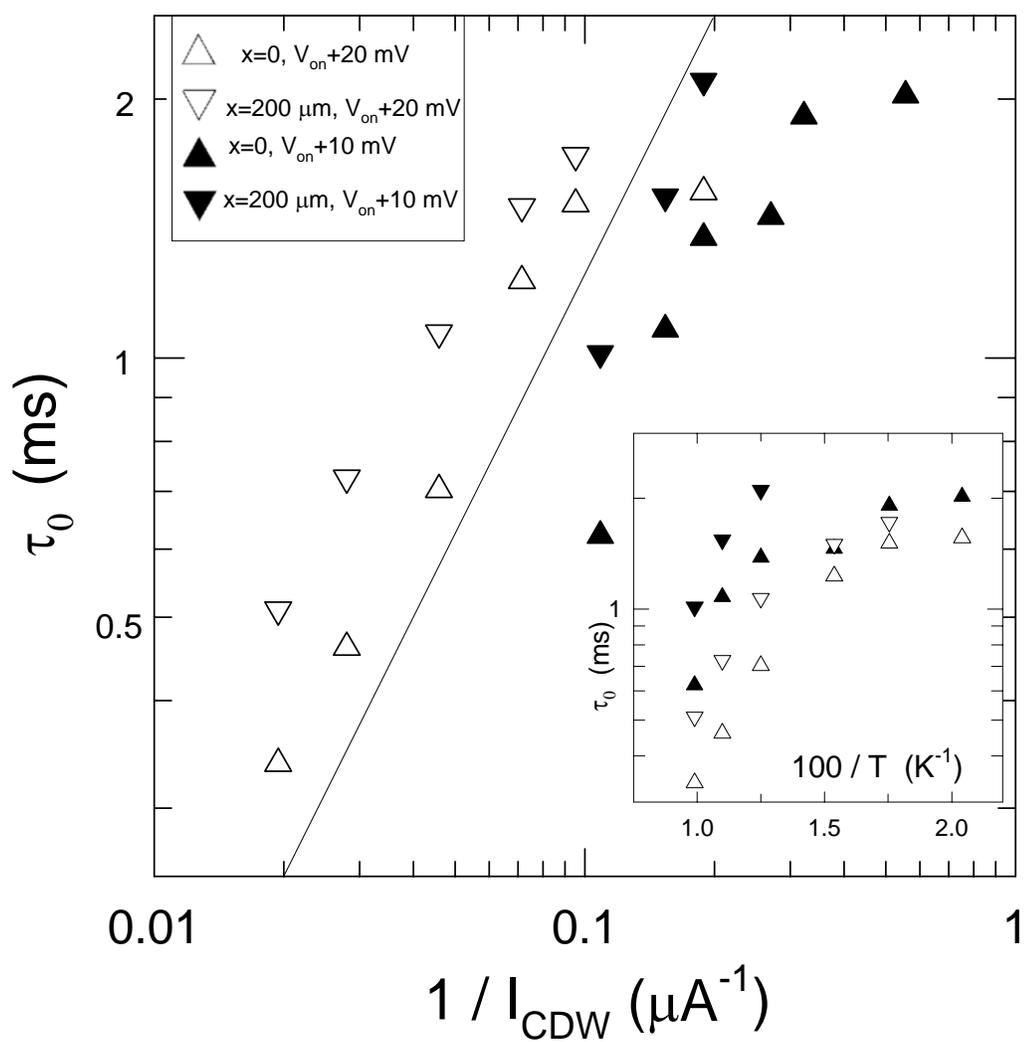

**Figure 11.** Relaxation times for *sample 4* ($\nu = 775$ cm$^{-1}$) plotted as functions of the temperature dependent CDW current at $V = V_{on} + 10$ mV and $V_{on} + 20$ mV. A line with slope = 1 is shown for reference. *Inset:* Temperature dependence of the relaxation times.

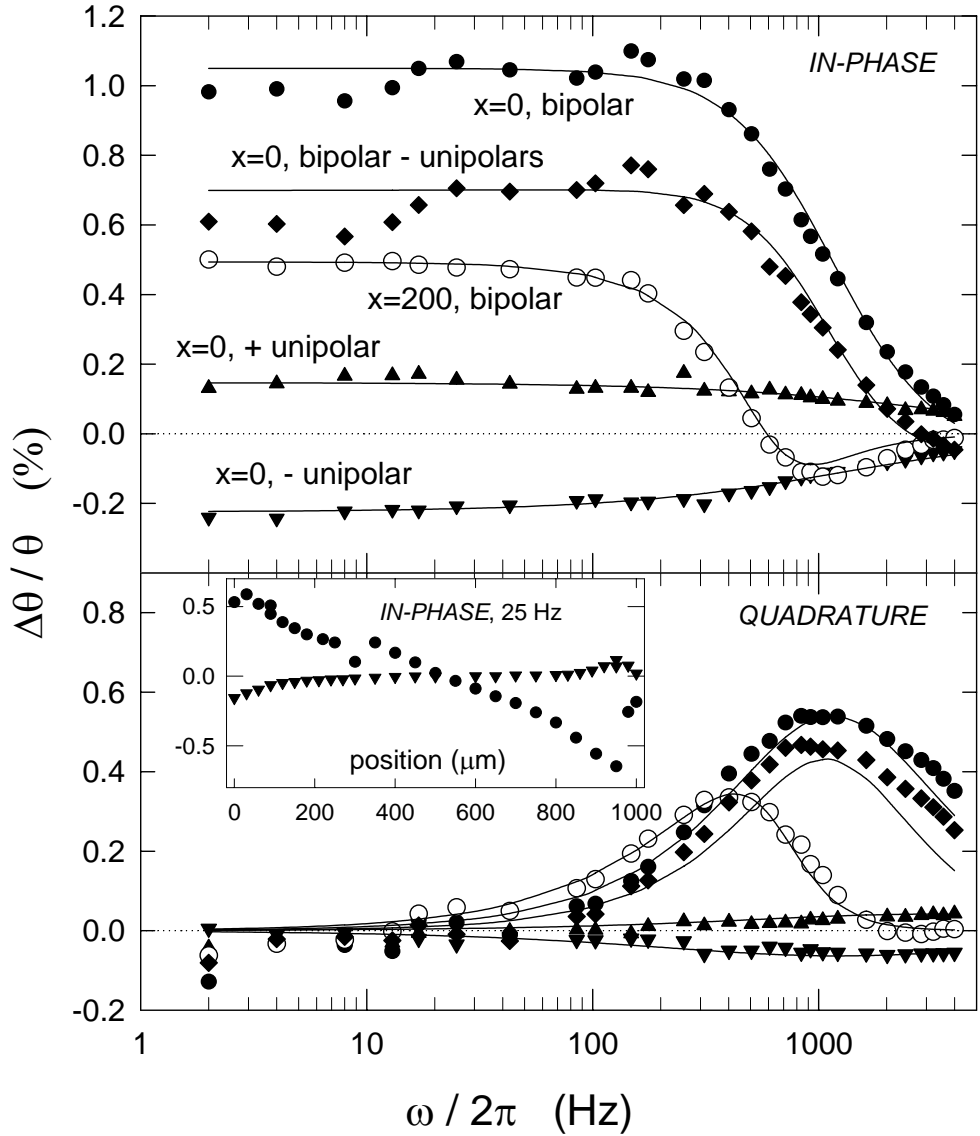

**Figure 12.** The frequency dependence of the electro-transmittance ($\nu = 820$ cm$^{-1}$) of *sample 3* at $V = 2\,V_T = 3.1\,V_{on}$ and $T = 101$ K. Shown are the measured in-phase and quadrature responses for bipolar square-waves at $x = 0$ and $x = 200\,\mu$m and positive and negative unipolar square-waves at $x = 0$. Also shown is the difference between the $x = 0$ bipolar and (difference of the) unipolar responses. The curves are fits to Eqtn. 1. *Inset*: The spatial dependence of the (in-phase) bipolar ($V = 1.5\,V_T$) and negative unipolar ($V = 2\,V_T$) responses at 25 Hz, for which the quadrature responses are negligible.

37